\documentclass[prl,twocolumn,superscriptaddress]{revtex4-1}

\usepackage{graphicx}

\begin{document}

\title{Ultrafast melting of a charge-density wave in the Mott insulator $1T$-TaS$_2$}

\author{S.~Hellmann}
\affiliation{Institute for Experimental and Applied Physics, University of Kiel, 24098 Kiel, Germany}
\author{M.~Beye}
\affiliation{Institute for Experimental Physics and Centre for Free-Electron Laser Science, University of Hamburg, 22761 Hamburg, Germany}
\author{C.~Sohrt}
\affiliation{Institute for Experimental and Applied Physics, University of Kiel, 24098 Kiel, Germany}
\author{T.~Rohwer}
\affiliation{Institute for Experimental and Applied Physics, University of Kiel, 24098 Kiel, Germany}
\author{F.~Sorgenfrei}
\affiliation{Institute for Experimental Physics and Centre for Free-Electron Laser Science, University of Hamburg, 22761 Hamburg, Germany}
\author{H.~Redlin}
\affiliation{Deutsches Elektronen-Synchrotron, DESY, 22607 Hamburg, Germany}

\author{M.~Kall\"ane}
\affiliation{Institute for Experimental and Applied Physics, University of Kiel, 24098 Kiel, Germany}
\author{M.~Marczynski-B\"uhlow}
\affiliation{Institute for Experimental and Applied Physics, University of Kiel, 24098 Kiel, Germany}
\author{F.~Hennies}
\affiliation{Institute for Experimental Physics and Centre for Free-Electron Laser Science, University of Hamburg, 22761 Hamburg, Germany}

\author{M.~Bauer}
\affiliation{Institute for Experimental and Applied Physics, University of Kiel, 24098 Kiel, Germany}
\author{A.~F\"ohlisch}
\altaffiliation[Present address: ]{Helmholtz-Zentrum Berlin f\"ur Materialien und Energie and Universit\"at Potsdam.}
\affiliation{Institute for Experimental Physics and Centre for Free-Electron Laser Science, University of 
Hamburg, 22761 Hamburg, Germany}
\author{L.~Kipp}
\affiliation{Institute for Experimental and Applied Physics, University of Kiel, 24098 Kiel, Germany}
\author{W.~Wurth}
\affiliation{Institute for Experimental Physics and Centre for Free-Electron Laser Science, University of Hamburg, 22761 Hamburg, Germany}
\author{K.~Rossnagel}
\email{rossnagel@physik.uni-kiel.de}
\affiliation{Institute for Experimental and Applied Physics, University of Kiel, 24098 Kiel, Germany}

\date{\today}             

\begin{abstract}
Femtosecond time-resolved core-level photoemission spectroscopy with a free-electron laser is used to measure the atomic-site specific charge-order dynamics in the charge-density-wave/Mott insulator $1T$-TaS$_2$. After strong photoexcitation, a prompt loss of charge order and subsequent fast equilibration dynamics of the electron-lattice system are observed. On the time scale of electron-phonon thermalization, about 1~ps, the system is driven across a phase transition from a long-range charge ordered state to a quasi-equilibrium state with domain-like short-range charge and lattice order. The experiment opens the way to study the nonequilibrium dynamics of condensed matter systems with full elemental, chemical, and atomic site selectivity.
\end{abstract}


\maketitle


Femtosecond time-resolved spectroscopy has become a powerful tool in condensed matter research because it delivers direct dynamical information at the fundamental time scale of elementary electronic processes \cite{CaMuUp_07,MiSaMa_08}. The method is particularly useful for complex materials, in which two or more of the lattice, charge, spin, and orbital degrees of freedom are strongly coupled. It allows to determine the nature and strength of the interactions between the degrees of freedom, to identify the dominant interactions, and thus to gain important insights into ground-state properties, thermally driven phase transitions, and, possibly, novel hidden phases \cite{HiPrTr_06,PeLoLi_06,ScKiBo_08,ToScSt_09}.

An exceedingly fertile ground for the combination of spectral selectivity and femtosecond time resolution has been found in materials, in which the charge and lattice degrees of freedom interact strongly to form a coupled charge-density-wave (CDW)/periodic-lattice-distortion (PLD) ground state. The quasiparticle and collective mode dynamics of the CDW/PLD state, the finite electron-lattice coupling time, and the collapse of the electronic gap under strong excitation are now known \cite{DeBiMi_99,DeFoBe_02,PeLoLi_06,PeLoLi_08,ScKiBo_08,ToScSt_09}. Yet, direct dynamical information on the CDW itself, i.e., on the local charge order, is missing. Specifically, it is not clear how fast a CDW can melt and recondense after impulsive excitation. The present study provides this fundamental piece of information for the layered reference compound $1T$-TaS$_2$ \cite{FazTos_79,DeFoBe_02,PeLoLi_06,PeLoLi_08,FrKrGe_09}.

Employing time-resolved x-ray photoemission spectroscopy (TR-XPS) with a free-electron laser \cite{Ackermann_07,PiFoBe_08}, we directly measure the melting of a large-amplitude CDW in $1T$-TaS$_2$ at atomic sites in real time. Our results show that long-range charge order collapses promptly after strong optical excitation and that a domain-like quasi-equilibrium CDW/PLD state is reached with a sub-picosecond time constant.

\begin{figure}[t]
\includegraphics[width=0.95\linewidth]{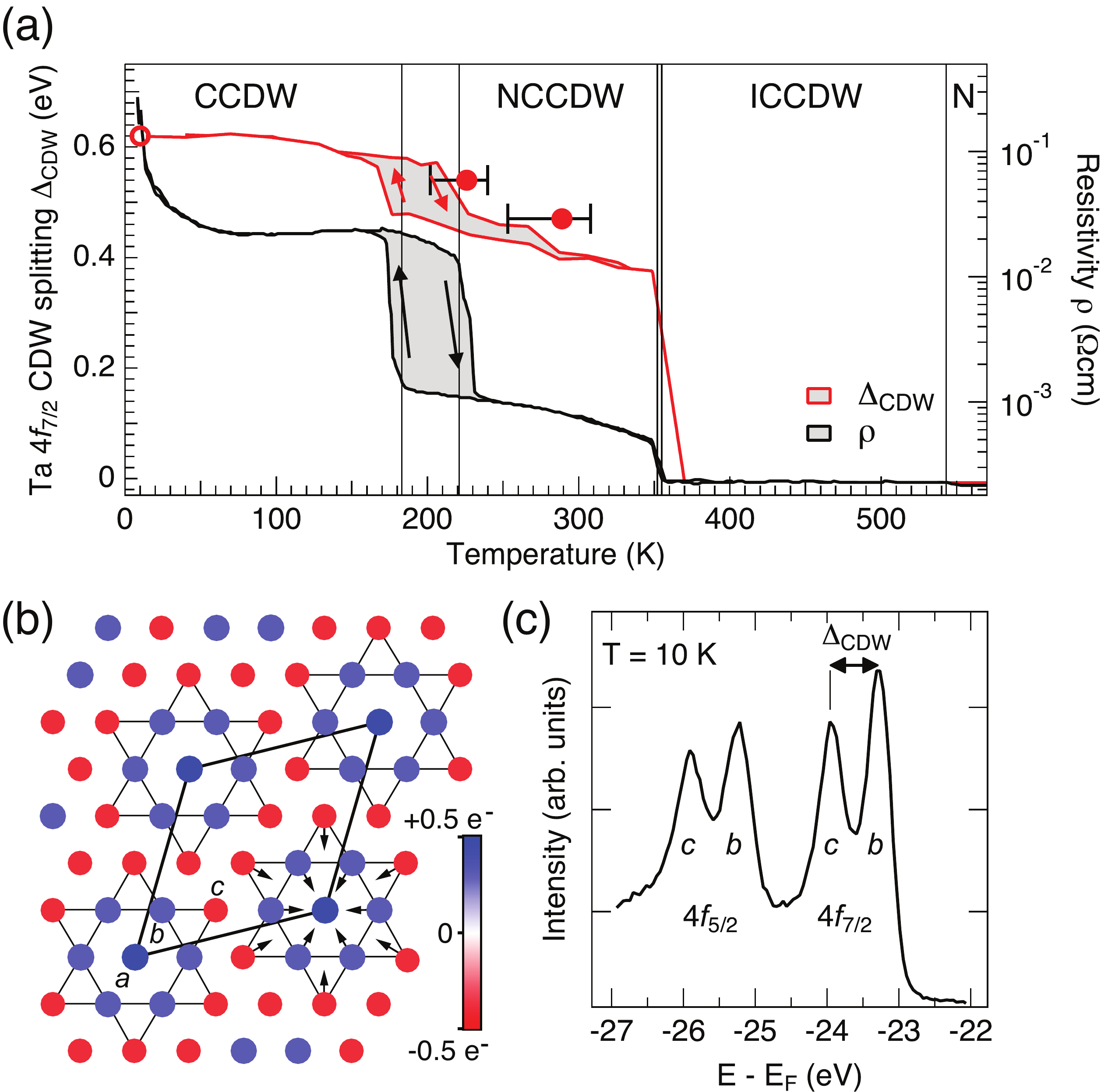}
\caption{\label{fig1} (color online).
Charge-density-wave phases of $1T$-TaS$_2$.
(a) Phase diagram with the normal undistorted phase (N) as well as incommensurate (IC), nearly commensurate (NC), and commensurate (C) charge-density-wave (CDW) phases. Temperature dependencies of the electrical resistivity (black line, Ref.~\onlinecite{SiKuAk_08}) and of the CDW-induced Ta $4f$ core-level splitting $\Delta_{CDW}$ (red line, Ref.~\onlinecite{HugPol_76}, values scaled to match our criterion for $\Delta_{CDW}$ \cite{dataanalysis}) are included. The initial and final $\Delta_{CDW}$ values of our pump-probe experiments are indicated by open and filled red circles, respectively.
(b) Sketch of the CCDW showing David-star clusters with inequivalent $a$, $b$, and $c$ Ta atoms. The arrows indicate the displacement of the Ta atoms from their original positions. The electron density increases towards the center of the cluster. 
(c) Ta $4f$ photoemission spectrum measured with a photon energy of $156$~eV. Each Ta $4f$ level is split into two peaks associated with sites $b$ and $c$, separated by $\Delta_{CDW}$. 
}
\end{figure}

The model system $1T$-TaS$_2$ is a complex material with a simple basic structure consisting of S--Ta--S sandwiches in which each Ta atom is octahedrally coordinated by six S atoms. The interesting physics in this compound is restricted to the hexagonal Ta layers and results from simultaneously strong electron-phonon and electron-electron interactions. The phase diagram includes a high-temperature metallic phase, incommensurate CDW and nearly commensurate CDW (NCCDW) phases at intermediate temperatures, and ultimately a low-temperature commensurate CDW (CCDW) phase coexisting with a Mott insulator phase [Fig.~\ref{fig1}(a)] \cite{FazTos_79,BaGhGu_84}.

Our focus here is on the CCDW-to-NCCDW transition. In the CCDW phase, the Ta atoms are grouped into David-star clusters consisting of two 6-atom rings which contract towards the central atom \cite{FazTos_79}. The Ta atoms are displaced by up to $7$\% of the in-plane lattice constant \cite{BroJel_80} and the accompanying $\sqrt{13} \times \sqrt{13}$ CDW also has a large amplitude, as roughly $0.4$ electrons are transferred from each atom in the outer ring to the inner atoms [Fig.~\ref{fig1}(b)] \cite{SmKeDi_85}. In momentum space, there is not one uniform CDW gap opening up at the Fermi level $E_F$. Rather, the electronic states in the partially filled Ta $5d$ band are regrouped into submanifolds \cite{SmKeDi_85} and the combination of spin-orbit coupling and CDW reconstruction leads to a distinct and narrow band at $E_F$ that is susceptible to a Mott-Hubbard transition \cite{RosSmi_06}. Upon going from the CCDW to the NCCDW phase, the important effect is loss of phase coherence. A domain superstructure emerges \cite{IshSat_91}, where the structure within the domains is that of the CCDW phase and the domain walls, or discommensurations, form a supposedly metallic network that suppresses the Mott phase \cite{SiKuAk_08}.

To study the local dynamics of the charge degree of freedom across the CCDW-NCCDW transition, one needs a local probe sensitive to the amplitude of the CDW as well as to its degree of commensurability and phase coherence. XPS is such a probe, as the spectra of shallow core levels generally reflect the charge distribution at atomic sites. In $1T$-TaS$_2$ it is the Ta $4f$ binding energy that is particularly sensitive to the local charge density. In the CCDW phase, there are three inequivalent Ta atoms in the unit cell designated as $a$, $b$, and $c$ in the ratio 1:6:6 [Fig.~\ref{fig1}(b)] so that, as shown in Fig.~\ref{fig1}(c), each Ta $4f$ level is split into two well separated peaks $b$ and $c$ (the $a$ peak is weak and not readily resolved \cite{HugSca_95}). In this phase the $b$--$c$ splitting $\Delta_{CDW}$ is a direct measure of the CDW amplitude \cite{HugPol_76}. In the NCCDW phase, on the other hand, the CDW-induced splitting is much less pronounced due to the broken phase coherence. Here $\Delta_{CDW}$ predominantly reflects the size of the commensurate domains relative to the area of the discommensurations \cite{ZwBeVo_98}. Figure~\ref{fig1}(a) demonstrates that $\Delta_{CDW}$ can indeed be regarded as an order parameter for the system, as the temperature dependence of $\Delta_{CDW}$ \cite{HugPol_76} mimicks the resistivity curve \cite{InOnTa_83,SiKuAk_08}.

Our TR-XPS measurements of the order parameter $\Delta_{CDW}$ were performed at the plane grating monochromator beamline PG2 \cite{MaWeHo_06,WeMaWu_07} of the free-electron laser at Hamburg (FLASH) \cite{Ackermann_07} in combination with a synchronized optical pump laser \cite{PiFoBe_08,GaAzBe_08,AzDuRa_09}. The $1T$-TaS$_2$ single crystals \cite{crystalgrowth} were excited by a $120$~fs (FWHM) pump pulse with a photon energy of $1.55$~eV. The photoelectrons used to probe the state of the system were generated by a delayed FLASH pulse of $\sim$$156$~eV, corresponding to the third harmonic of the fundamental. The total energy and time resolutions, including instrumental resolutions, space-charge broadening \cite{HeRoMa_09}, and temporal jitter \cite{AzDuRa_09}, were $300$~meV (FWHM) and $700$~fs (FWHM), respectively. All measured kinetic energies were referenced to the Fermi edge of the unpumped system. Starting with $1T$-TaS$_2$ deep in the CCDW phase ($T=10$~K), two optical excitation fluences were used, $F_1=1.8$~mJ/cm$^2$ and $F_2=2.5$~mJ/cm$^2$, corresponding to absorbed energy densities of $\sim$$120$~meV/Ta and $\sim$$165$~meV/Ta, respectively. These values are close to the electronic band energy gain in the CCDW phase ($\sim$$160$~meV/Ta \cite{RosSmi_06}) and they are larger than the energy density required to heat the excited volume to above the CCDW-NCCDW transition temperature ($\sim$$110$~meV/Ta). We therefore expect photoinduced transitions in the course of which charge {\it and} lattice order are melting.

\begin{figure}[t]
\includegraphics[width=0.75\linewidth]{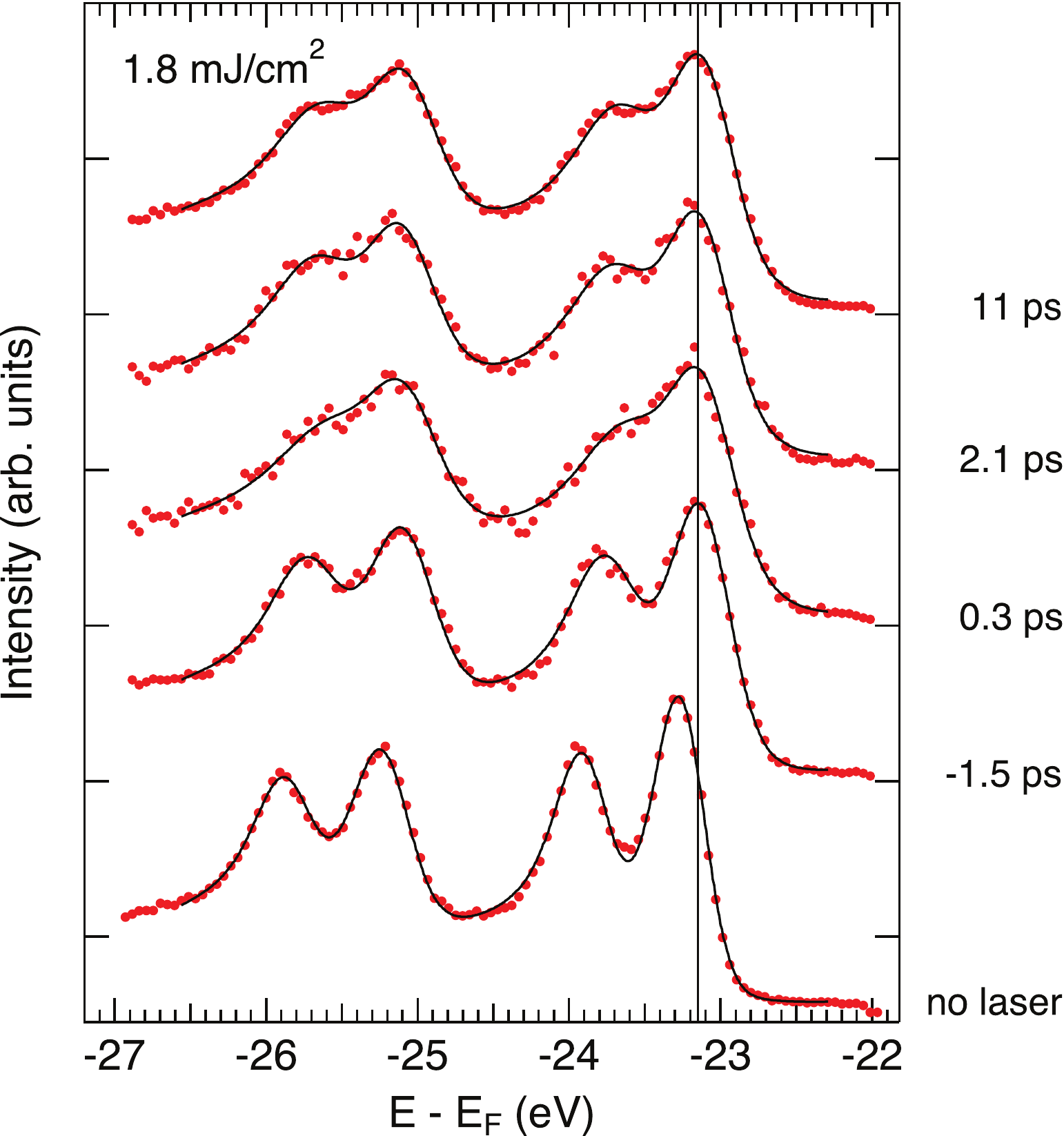}
\caption{\label{fig2} (color online).
Time-resolved Ta $4f$ photoemission spectra of $1T$-TaS$_2$ for  selected pump-probe delays. Red dots represent the experimental data, solid black curves are fits to the spectra \cite{dataanalysis}, and the vertical line is a guide to the eye illustrating the pump laser induced space-charge shift.}
\end{figure}

Figure~\ref{fig2} shows selected time-resolved Ta $4f$ core-level spectra. Two processes governing the response of the system to the impulsive optical excitation can be readily identified: (i) a sub-picosecond reduction of the CDW-induced Ta $4f$ splitting, followed (ii) by a partial recovery on the picosecond time scale into a quasi-equilibrium state with a lifetime of longer than 10~ps. It should be emphasized that both processes are much faster than, and therefore well separated from, the dynamics of the spectral shift and broadening due to the vacuum space charge created by the pump laser (compare the positions of the leading peak in the top four spectra and in the bottom spectrum in Fig.~\ref{fig2}).

\begin{figure}[t!]
\includegraphics[width=0.83\linewidth]{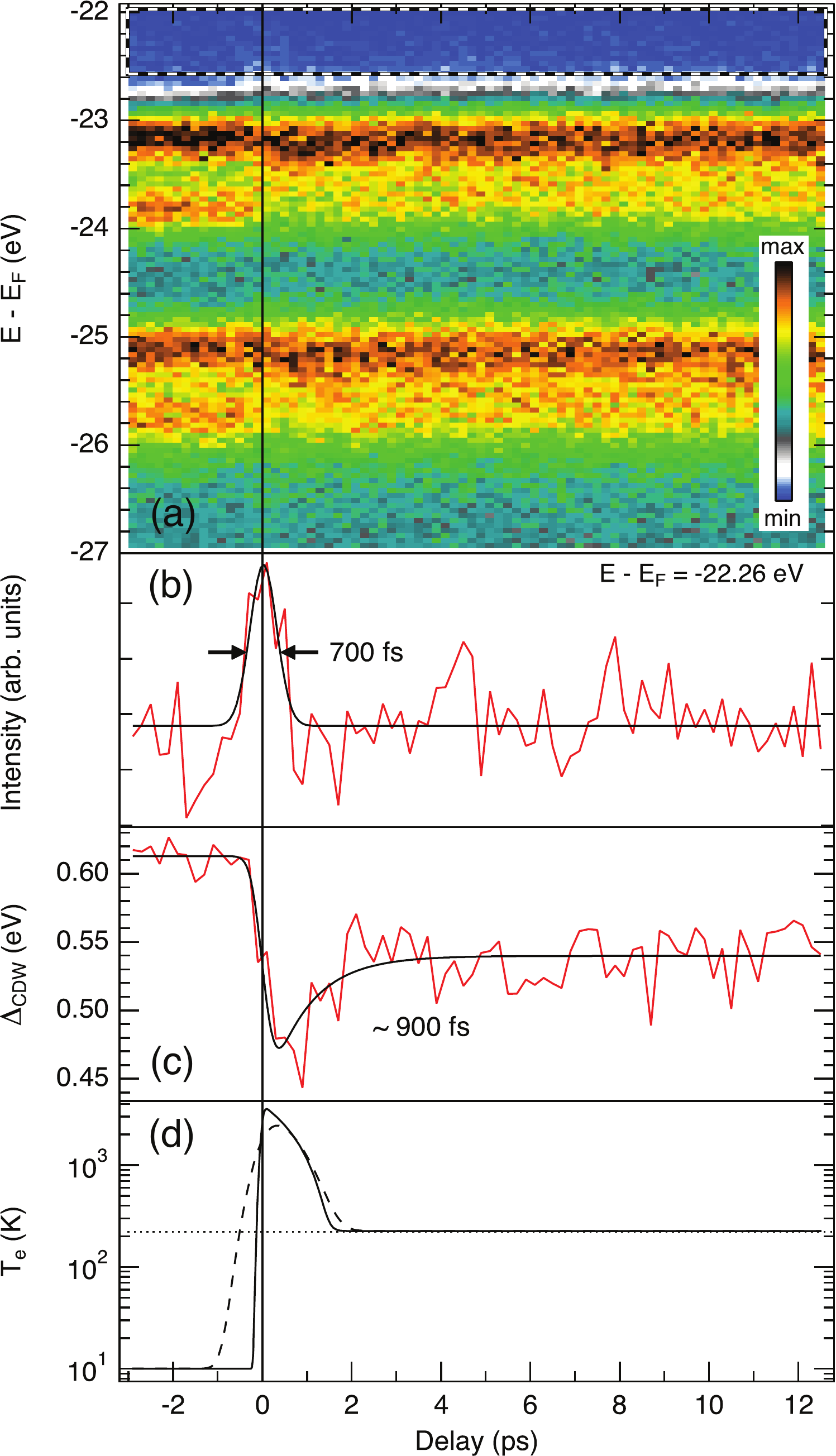}
\caption{\label{fig3} (color online).
Charge-order dynamics in $1T$-TaS$_2$.
(a) Time-resolved Ta $4f$ photoemission spectra of $1T$-TaS$_2$ as a function of pump-probe delay, measured at a pump fluence of $1.8$~mJ/cm$^2$. The dashed box centered on $-22.26$~eV indicates where first-order sideband intensity of the $c$ peak of the Ta $4f_{7/2}$ level is to be expected.
(b) Photoemission intensity integrated over the energy interval marked in (a), representing the cross-correlation between pump and probe pulses (red curve). The black curve is a Gaussian fit.
(c) Ta $4f$ splitting $\Delta_{CDW}$ as a function of pump-probe delay (red curve), obtained by line shape fitting \cite{dataanalysis}. The black curve is a fit to the red curve using a single exponential starting at $t=0$ convoluted with a Gaussian representing the effective time resolution.
(d) Electron temperature $T_e$ as a function of pump-probe delay (solid curve), as calculated from the two-temperature model for the parameters of the experiment \cite{parameters}. Accounting for the experimental time resolution leads to the dashed curve. The dotted horizontal line marks the equilibrium CCDW-NCCDW transition temperature.
}
\end{figure}

The ultrafast vaporization and equilibration dynamics shall now be exposed in more detail by analyzing the photoemission intensity map of Fig.~\ref{fig3}(a), measured at the lower pump fluence $F_1$. Note that the effective time resolution of the experiment [$700$~fs (FWHM)] is directly reflected in the time dependence of the (first-order) Ta $4f_{7/2}$ sideband intensity [Fig.~\ref{fig3}(b)], which arises from the absorption of optical laser photons during the ionization process (laser assisted photoelectric effect) \cite{GlScCh_96}.

The ultrafast dynamics of the CDW-induced Ta $4f$ splitting is brought to light in Fig.~\ref{fig3}(c) which displays the fitted $\Delta_{CDW}$ as a function of the pump-probe delay \cite{dataanalysis}. Within the time resolution of the experiment, $\Delta_{CDW}$ drops from the original value of $0.62$~eV to $0.47$~eV, and it then recovers partially to a quasi-equlibrium value $\Delta_{CDW,1}^\ast = 0.54$~eV with a time constant of $\sim$$900$~fs. In contrast to recent valence-band photo\-emission measurements in the moderate excitation regime \cite{PeLoLi_06,PeLoLi_08}, a periodic modulation of the time-resolved signal cannot be observed. Such modulation is generally caused by the CDW amplitude mode, but is seen to be strongly suppressed in the strong excitation regime, as in our case, because of scattering of the amplitude mode with a high density of excited quasiparticles \cite{PeLoLi_06,PeLoLi_08,ScKiBo_08,ToScSt_09}. We note that for the higher fluence $F_2$ the quasi-equilibrium $\Delta_{CDW}$ value, reached after the initial drop and recovery, is $\Delta_{CDW,2}^\ast = 0.47$~eV.

The corresponding quasi-equilibrium temperature of the electron-lattice system, $T^\ast$, can be estimated within the two-temperature model \cite{AnKaPe_74}. For the parameters of our experiment \cite{parameters}, we obtain $T_1^\ast = 226^{+14}_{-24}$~K and $T_2^\ast = 289^{+19}_{-36}$~K for the two pump fluences used. When plotted in the phase diagram of Fig.~\ref{fig1}(a) (red circles), the two photoinduced quasi-equilibrium states $(T_i^\ast, \Delta_{CDW,i}^\ast)$ fall, within the error bars, on the equilibrium $\Delta_{CDW}(T)$ curve in the region of the NCCDW phase, thus indicating ultrafast transitions to CDW/PLD states with smaller distortion amplitude and broken phase coherence. For the lower fluence $F_1$, the calculated time dependence of the electron temperature is shown in Fig.~\ref{fig3}(d). The clear anti-correlation with the measured $\Delta_{CDW}$ [Fig.~\ref{fig3}(c)] suggests that the charge order parameter is tied to the electron temperature.

We conclude that the observed response of the Ta $4f$ CDW splitting provides strong evidence for the two anticipated melting processes. The impulsive optical excitation causes first a quasi-instantaneous collapse of the charge order by rapidly heating the electrons to a highly elevated temperature and then, when the electrons cool down by transferring their energy to the lattice, a melting of the long-range lattice order of the CCDW phase. This scenario implies that the CDW and the PLD, which are so strongly coupled in equilibrium that one cannot exist without the other \cite{JohMaz_08}, are decoupled after photoexcitation on the time scale for electron-phonon thermalization \cite{ToScSt_09}.

Two more implications of our results are noteworthy. First, the transient collapse of charge order happens on the same time scale as the collapse of the Mott-Hubbard gap observed in time-resolved valence-band measurements \cite{PeLoLi_06,PeLoLi_08}. This suggests that the loss of order in the electronic portion of the commensurate CDW/PLD may be a factor in the ultrafast melting of the Mott phase. Second, the picosecond equilibration time indicates that the CDW/PLD does not relax into the CCDW domain superstructure of the equilibrium NCCDW phase, which would require the ultrafast nucleation and growth of a (metallic) discommensuration network. It seems more likely that the system relaxes into a state with the same relative size  of commensurate and distorted areas, but rather with an inhomogeneous distribution of distorted (metallic) islands in a CCDW background.

In conclusion, strong photoexcitation of the CDW/Mott insulator $1T$-TaS$_2$ leads to an ultrafast, temporally decoupled melting of long-range commensurate charge and lattice order: After the prompt collapse of the charge order, the electron-lattice system rapidly equilibrates into a domain-like short-range ordered CDW/PLD phase on the picosecond time scale of electron-phonon thermalization. This result constitutes the first direct measurement of charge-order dynamics in a complex material with combined femtosecond resolution and atomic site sensitivity, thereby establishing the technique of TR-XPS at a free-electron laser. As the time and energy resolution of this element, chemically, and atomic-site selective technique are improving, we can expect many novel insights into the nonequilibrium behavior of condensed matter systems. TR-XPS specifically enables us to create detailed movies of local electron dynamics, not only of ultrafast phase transitions in complex materials but also of chemical reactions on solid surfaces.

We thank T.~Beeck, M.~Berglund, S.~D\"usterer, B.~Fominykh, S.~Gieschen, N.~Guerassimova, S.~Harm, J.~T.~Hoeft, V.~Kocharyan, M.~Kuhlmann, S.~Lang, K.~Malessa, H.~Meyer, A.~Pietzsch, T.~Riedel, V.~Rybnikov, W.~Schlotter, N.~Stojanovic, C.~Thede, R.~Treusch, M.~Wellh\"ofer, and the FLASH operators for experimental support at various stages of this project.
Financial support by the German Federal Ministry of Education and Research through the priority program FSP 301 {\it FLASH: Matter in light of ultrashort and extremely intense x-ray pulses} and by the DFG graduate school 1355 {\it Physics with new advanced coherent radiation sources} is gratefully acknowledged.



\begin{thebibliography}{99}
\bibitem{crystalgrowth}
$1T$-TaS$_2$ single crystals were grown in a temperature gradient of $840$--$870^\circ$C in 1000~h using iodine vapor transport with excess sulphur ($2.5$~mg/cm$^3$) \cite{InOnTa_83}.
\bibitem{dataanalysis}
The model function consists of four Doniach-{\u S}unji{\'c} lines (plus a linear background) convoluted with a Gaussian. In the fits the Gaussian width from the instrumental resolution, the Ta $4f$ spin-orbit splitting, and the asymmetry parameters were constrained to the known values \cite{HugSca_95}. The fitted charge-order parameter $\Delta_{CDW}$ represents the CDW-induced energy separation between the two Ta $4f$ spin-orbit doublets [one parameter for the two $b$--$c$ splittings in Fig.~\ref{fig1}(a)].
\bibitem{parameters}
The material specific parameters are: electronic heat capacity coefficient $\gamma = 40.3$~Jm$^{-3}$K$^{-2}$ \cite{Wilson_75}, Debye temperature $\Theta_D = 237$~K \cite{Schwall_76}, phonon heat capacity coefficient $\beta=12.6$~Jm$^{-3}$K$^{-4}$ \cite{Schwall_76}, mass enhancement parameter $\lambda = 0.85$, reflectivity $R = 0.39$ \cite{Beal_75}, and absorption coefficient $\alpha=3\cdot 10^7$ m$^{-1}$ \cite{Beal_75}.
\end{thebibliography}
\end{document}